\documentclass[useAMS, usenatbib, usegraphicx]{mn2e}
\usepackage{amssymb}
\usepackage{booktabs}
\newcommand\ApJ{ApJ}
\newcommand\MNRAS{MNRAS}

\newcommand\PASJ{PASJ}

\newcommand\ApJS{ApJS}
\newcommand\ApSS{Ap\&SS}
\newcommand\AnA{A\&A}

\def\alf{Alfv\'en\,}
\def\alfc{Alfv\'enic\,}
\def\sg{\sigma}
\def\va{v_A}

\def\bq{\begin{equation}}
\def\eq{\end{equation}}
\def\ee #1 {\times 10^{#1}}
\def\ut #1 #2 { \, \rmn{#1}^{#2}}
\def\u #1 { \, \rmn{#1}}

\let\grad=\nabla
\def\B{{\bf{B}}}
\def\v{{\bf{v}}}
\def\vB{\bmath{v}_B}
\def\hB{\hat{{\bf{B}}}}

\newcommand\cross{\bmath{\times}}
\def\curl{{\grad \cross}}
\def\div #1 {\grad \cdot #1}

\def\u{\bmath{u}} 
\def\J{\bmath{J}}
\def\Jpa{\bmath{J_\parallel}}  % J_||

\newcommand{\delt} [1] {\frac{\partial #1}{\partial t}}

\title{Can Hall effect trigger Kelvin-—Helmholtz instability in sub--\alfc flows?}
\author[B.P.Pandey]
        {B.P. Pandey\\
{Department of Physics \& Astronomy, Macquarie University, Sydney, NSW 2109, Australia} }
\date{\today}
\pagerange{\pageref{firstpage}--\pageref{lastpage}}
\pubyear{2014}
\begin{document}
\maketitle
\label{firstpage}
\begin{abstract}
In the Hall magnetohydrodynamics, the onset condition of the Kelvin—-Helmholtz instability is solely determined by the Hall effect and is independent of the nature of shear flows. In addition, the physical mechanism behind the super and sub--\alfc flows becoming unstable is quite different: the high frequency right circularly polarized whistler becomes unstable in the super--\alfc flows whereas low frequency, left circularly polarized ion-—cyclotron wave becomes unstable in the presence of sub--\alfc shear flows. The growth rate of the Kelvin—-Helmholtz instability in the super--\alfc case is higher than the corresponding ideal magnetohydrodynamic rate. In the sub--\alfc case, the Hall effect opens up a new, hitherto inaccessible (to the magnetohydrodynamics) channel through which the partially or fully ionized fluid can become Kelvin-—Helmholtz unstable. The instability growth rate in this case is smaller than the super--\alfc case owing to the smaller free shear energy content of the flow. 

When the Hall term is somewhat smaller than the advection term in the induction equation, the Hall effect is also responsible for the appearance of a new overstable mode whose growth rate is smaller than the purely growing Kelvin—-Helmholtz mode. On the other hand, when the Hall diffusion dominates the advection term, the growth rate of the instability depends only on the \alf-—Mach number and is independent of the Hall diffusion coefficient. Further, the growth rate in this case linearly increase with the \alf frequency with smaller slope for sub--\alfc flows. 
\end{abstract}


\section{Introduction}
The Kelvin—-Helmholtz (KH) instability is a macroscopic instability that grows at the boundary of the velocity shear layer in the fluid. Onset of the KH instability in an ideal magnetohydrodynamic fluid depends on an interplay between the shear flow stress and the magnetic tension force. In the presence of magnetic field when the shear flow speed is larger than the \alf speed, i.e. when the flow is super--\alfc, the medium becomes Kelvin—-Helmholtz (KH) unstable.  The instability can be completely quenched if the flow is sub--\alfc i.e. when the flow speed is smaller than the \alf speed. A typical example where KH instability may occur is when the fluid flows at two different velocities parallel to the surface of discontinuity. If the shear flow can overcome the surface tension, or the magnetic tension force, the fluid becomes unstable \citep{C61}.  

Given the prevalence of shear flows in the geo, space and astrophysical environment, it is not surprising that the KH instability is invoked as an efficient mechanism for the transport of momentum and energy across the shear layer. For example, KH instability is one of the most important mechanism for the plasma transport across Earth{}\'s magnetosphere \citep{M84, M87, FT91}. The observation of the vortices at the surface of the coronal mass ejecta hints at the likely presence of KH instability at the Sun \citep{F11}. Recent observations of the quiescent prominences in the solar atmosphere suggests the presence of turbulent flows may have been caused by the KH instability \citep{M15}. The Kelvin—-Helmholtz eddies may drive turbulence and heat ambient plasma in the solar atmosphere \citep{K15, M16}. Far away ($\gtrsim 100\,\mbox{AU}$) from the sun, when the hot solar wind meets the cold interstellar medium, the boundary layer between them may become Kelvin--Helmholtz unstable \citep{D11}. Shear instabilities in the dust layer of the solar nebula may inhibit the formation of planetisimals \citep{S98, SI01, MI06, HT14}. It is quite plausible that the KH instability is responsible for the narrower line profiles in the ionized spices in molecular clouds \citep{WZ04}. Therefore, from the wind driven parched plains of the Earth to the vast expanse of the interstellar medium, Kelvin—-Helmholtz instability plays an important role in driving the efficient turbulent mixing of the mass and momentum in the medium. 

The onset condition of the KH instability in a magnetized ideal fluid is determined by the \alf-—Mach number, defined as 
\bq
M_A = \left(\frac{v}{\va}\right)\,,
\eq
where $v = |\v|$ is the shear--flow speed and $\va = B/\sqrt{4\,\pi\,\rho}$ is the \alf speed with $B$ and $\rho$ as the magnetic field and the mass density of the fluid respectively. In the presence of a magnetic field  if the force due to the shear flow stress is larger than the magnetic tension force, i.e. when the flow is super--\alfc ($M_A > 1$), the fluid becomes unstable. Opposite is the case when the magnetic tension force dominates the shear flow stress. These conclusions are valid of course in the absence of any additional source of free energy. However, if the magnetized fluid has an additional source of free energy such as the flow aligned uniform current (manifested as the twist in the flow aligned magnetic field) \citep{Z10, Z14, Z15}, or, in planar jets \citep{ST93}, the sub--\alfc ideal magnetohydrodynamic (MHD) flows may also become KH unstable. In the absence of any such additional free energy source however, only super--\alfc flows are unstable in the single fluid ideal MHD framework \citep{C61}. 

Note that in the ideal MHD framework, the bulk fluid is magnetized and thus the fluid as a whole experiences the magnetic tension force. However, in the Hall MHD since the ion cyclotron frequency is much less than the dynamical frequency of interest [which in a partially ionized plasma is the neutral--ion collision frequency \citep{PW06, PW08}] the ions are very weakly magnetized and thus are incapable of directly experiencing the magnetic tension force. It is the magnetized electrons that experiences the magnetic force and thus, the role of magnetic tension force in mitigating the instability is not guaranteed in the presence of Hall effect.    

For the MHD description to be valid, the wavelength of fluctuations must be much larger than the Hall scale \citep{PW08}, a constraint not always satisfied in the geo and astrophysical environment. For example, the observations of the Earth{}\'s magnetopause boundary suggest that the Hall scale is comparable to the thickness of the boundary layer \citep{BR82, M02}. The origin of the whistler waves in the Earth{}\'s lower ionosphere is due to the Hall effect \citep{S99, A05, P16}. The Hall diffusion is important in the large part of the solar atmosphere \citep{PW12, SPH12, SPHC15}. In the protoplanetary discs, depending on the magnetic field strength, Hall diffusion could be important near the planet forming midplane region of the disc \citep{W07}. The Hall effect qualitatively changes the behaviour of the magnetic field in the molecular clouds \citep{W04}. Clearly, the Hall diffusion of the magnetic field could be crucial to the KH instability driven turbulence in diverse physical settings.\footnote{We shall use Hall effect and Hall diffusion terms interchangeably as they both mean one and the same thing}.       

The KH instability in the framework of Hall MHD for super--\alfc flows have been studied in the past both analytically \citep{TK65, SC68} (hereafter TK65, SC68) and numerically \citep{FT91, C03, JD11, JD12, H13}. The early analytical results suggested that in an incompressible flow, Hall effect not only causes faster growth of the instability in super--\alfc flows but even marginal ($M_A=1$) state becomes unstable. However, these early studies have been confined only to the super--\alfc or, marginal flows and do not consider the sub--\alfc case. The numerical investigation of the 2D incompressible parallel field--flow by \cite{C03} in agreement with the TK65 and SC68 finds that the KH instability is destabilized by the Hall effect. Recent simulation results of \cite{H13} also gives slightly larger (than the ideal MHD) growth rate in the Hall MHD.  However, the numerical simulations of the compressible super--\alfc ($M_A =2.5-5$) trans--sonic (flow speed of the order of or, larger than the sound speed) flows finds that the growth rate of the KH instability is unaffected by the Hall effect \citep{FT91}. The investigation of the KH instability in the partially ionized medium for the highly super--\alfc ($M_A=10$) trans--sonic flows also gives similar results \citep{JD11, JD12}. Clearly, the numerical investigation of the compressible and incompressible magnetized fluid gives different results. This is not surprising given that in a compressible fluid part of the free shear flow energy goes into the compression of the fluid thus slightly reducing the growth rate of the KH instability \citep{KG16}. To summarize the numerical investigation of the KH instability for both the compressible and incompressible flows in the framework of Hall MHD have been limited to the marginal or, super--\alfc flows. 

In the present work we investigate the Kelvin-—Helmholtz instability in the Hall MHD framework. Although the past investigations of the KH instability have been carried out for the fully ionized plasma, the same can be easily generalized to the partially ionized plasmas. Thus we shall use the basic set of non--ideal MHD equations for the partially ionized plasma given by \cite{PW06, PW08}. The advantage of such a formulation is that the results of the fully and weakly ionized plasmas constitute the subset of the general results and thus have wider applicability, ranging from the protoplanetary discs to the Earth{}\'s ionosphere. We briefly discuss the basic set of equations and boundary condition in section II. The detailed discussion of the boundary condition is given elsewhere \citep{R79}. We discuss the dispersion relation in section III. Since the Hall effect depends on the sign of the magnetic field,  we show that the KH instability growth rate is mirror symmetric with respect to such a sign change. The application of the result with a brief summary is given in section IV.
\section{Basic Model}
The basic set of non--ideal MHD equations which describes the partially ionized medium are \citep{PW06, PW08}
\bq
\delt \rho + \grad\cdot\left(\rho\,\v\right) = 0\,.
\label{eq:cont}
\eq
\bq
\rho\,\frac{d\v}{dt}=  - \nabla\,P + \frac{\J\cross\B}{c}\,.
\label{eq:meq}
\eq
\bq
\delt \B = \curl\left[
\left(\v + \v_B\right)\cross\B
- \frac {4\,\pi\,\eta_O}{c}\,\Jpa\right]\,.
\label{eq:indA}
\eq
\bq
\J = \frac{c}{4\,\pi}\curl\B\,.
\eq
Here 
\bq
\vB = \eta_P\,\frac{\left(\grad\cross\B\right)_{\perp}\cross\hB}{B} -– 
\eta_H\,\frac{\left(\grad\cross\B\right)_{\perp}}{B}\,,
\label{eq:md0}
\eq
is the magnetic diffusion velocity and $\eta_P = \eta_O + \eta_A$ is the Pedersen diffusivity. The parallel and perpendicular component of the current refers to its orientation with respect to the background magnetic field. 

The Ohm ($\eta_O$), ambipolar ($\eta_A$) and Hall ($\eta_H$) diffusivities are 
\bq \eta_O =
\frac{c^2}{4\,\pi\sigma}\,\,, \eta_{A} =
\frac{D^2\,B^2}{4\,\pi\,\rho_i\,\nu_{in}}\,, \eta_H =
\frac{c\,B}{4\,\pi\,e\,n_e}\,.
\label{eq:diffu}
 \eq
Here 
\bq
\sigma = \frac{c\,e\,n_e}{B}\, \left[\frac{\omega_{ce}}{\nu_{e}} + \frac{\omega_{ci}}{\nu_{i}} \right]
\eq 
is the parallel conductivity, $\omega_{cj} = e\,B/m_j\,c$ is particle$\textquoteright$s cyclotron frequency where $e \,,B\,,m_j\,,c$ denotes the charge, magnetic field, mass and speed of light respectively and  $D = \rho_n/\rho$ is the ratio of the neutral and bulk mass densities. For the electrons $\nu_{e} = \nu_{en}$ and for the ions $\nu_{i} \equiv \nu_{in}$. Although $\nu_{ee}\,,\nu_{ei}\,,\nu_{ii}\,\mbox{and}\,,\nu_{ie}$ can become comparable to  $\nu_{en}$ for example in the solar atmosphere \citep{PVK08, PW13}, it is the neutral-—plasma collision that gives rise to the ambipolar and Hall diffusion in the partially ionized medium. The collisions between the like particles $\nu_{ee},\,\nu_{ii}$ to the leading order do not cause any diffusion. The electron—-ion collision contributes to the Ohm diffusion. 

Defining plasma Hall parameter $\beta_j$ as
\bq
\beta_j = \frac{\omega_{cj}}{\nu_{jn}}\,,
\eq
and the Hall frequency
\bq
\omega_H  = \frac{\rho_i}{\rho}\,\omega_{ci} \approx X_e\,\omega_{ci}\,, 
\eq  
above diffsivities, Eq.~(\ref{eq:diffu}) can be written in the compact form \citep{PW08}
\bq
\eta_H = \left(\frac{v_A^2}{\omega_H}\right)\,,
\eta_A = D\,\left(\frac{v_A^2}{\nu_{ni}}\right)\,,
\mbox{and}\,,
\eta_O = \beta_e^{-1}\,\eta_H\,.
\label{eq:hdc}
\eq
Here $\rho_i = m_i\,n_i$ is the ion mass density and $X_e = n_e / n_n$ is the ratio of the electron ($n_e$) and neutral ($n_n$) number densities and is a measure of the fractional ionization of the medium. The collision frequency $\nu_{ni} = \rho_i\,\nu_{in} / \rho_n$.

The relative strength of various non-—ideal MHD effect is encapsulated in the plasma Hall parameter $\beta_j$. As is transparent from Eq.~(\ref{eq:hdc}), Hall dominates ambipolar when $\beta_i\ll1$ implying that the relative ion--neutral drift is unimportant in comparison to the electron—-ion drift.  Similarly, when the electrons are frozen in the partially ionized fluid, i.e. $\beta_e\gg1$, compared to the Hall the Ohm diffusion is negligible. We shall work in the limit $\beta_i\ll 1\ll \beta_e$ and thus neglect the ambipolar and Ohm diffusion and retain only the non-—ideal Hall term in the induction Eq.~(\ref{eq:indA}).    

Note that in the absence of Hall term, the induction Eq.~(\ref{eq:indA}) is invariant under the sign reversal of the magnetic field, i.e. $\B \rightarrow - \B$. However, the presence of Hall breaks this symmetry as the electrons and ions (and the neutrals if the plasma is partially ionized) become two uncoupled fluid with the unmagnetized ions (and neutrals) carrying the inertia of the fluid and the magnetized electrons carrying the current. It is well known that the sign change of the magnetic field affects the way angular momentum is transported in the Hall dominated region of the accretion discs \citep{W99}.   

Within the Hall scale \citep{PW08}
\bq
L_H = \frac{\va}{\omega_H}\,,
\eq  
the ions, unlike in the ideal MHD, are weakly coupled to the magnetic field (due to their frequent collision with the neutrals) and thus do not {\it feel} the magnetic tension force. As a result it is only the magnetized electrons that affects the onset of the KH instability in the super--\alfc flows when the Hall scale is greater than the fluctuation wavelength. The ions in this case provides only the neutralizing background. However, when the fluctuation wavelength is larger than the Hall scale, the weakly magnetized ions cause novel low frequency Kelvin-—Helmholtz mode in the sub--\alfc flows. This new mode is akin to the well--known electrostatic ion—-cyclotron instability in the fully ionized plasmas \citep{Ml02}.     

Defining following momentum flux and magnetic flux tensors
\begin{eqnarray}
\bf{\Pi} &=& p\,\bf{I} + \rho\,{\v\,\v} + \frac{1}{4\,\pi}\Big[\frac{\B^2}{2}\,\bf{I} - \left(\B\cdot\nabla\right)\B\Big] 
\nonumber\\
\bf{M} &=& \left(\v + \vB\right)\,\B - \B\,\left(\v + \vB\right)\,,
\end{eqnarray}
the momentum and inductions equations (\ref{eq:meq}) and Eqs.~(\ref{eq:indA}) can be written in the following conservative form
\bq
\delt{\left(\rho\,\v\right)} + \grad\cdot \bf{\Pi} = 0\,
\label{eq:meq1}
\eq
\bq
\delt \B + \grad\cdot \bf{M} = 0\,.
\label{eq:indB}
\eq
The surface of flow discontinuity requires that the certain conditions must be satisfied across the surface. Thus considering an element of the surface  in the rest frame of the fluid, from Eqs.~(\ref{eq:cont}) and Eq (\ref{eq:meq1}) one gets \citep{LL63}
\bq
\big[\rho\,v_n\big] = 0\,,
\label{eq:bc1}
\eq
\begin{eqnarray}
\big[p + \rho\,v_n^2 + \left(\B_t^2 - \B_n^2\right)/8\,\pi\big] &=& 0\,,
\nonumber\\  
\big[\rho\,v_n\,\v_t - \frac{B_n\,\B_t}{4\,\pi}\big]&=& 0\,.
\label{eq:bc2}
\end{eqnarray}
where the suffix $n$ and $t$ pertains to the components normal and tangent to the surface and [f] denotes the discontinuity in the $f$ across the surface. 

It is customary to use the Ohm{}\'s law and infer the final boundary condition from
\bq
\big[B_n] = 0\,,
\label{eq:bc3}
\eq 
and the continuity of the tangential component of the electric field \citep{LL63, R79}. Since the bulk fluid velocity $\v$ and the magnetic field $\B$ provides a direct description of the macroscopic behaviour of the medium in the MHD framework \citep{P07}, it is desirable to formulate the boundary conditions without any reference to the electric field. The necessity to think in terms of electric field near the surface of discontinuity probably emanates from our inability to put the Ohm{}\'s law in the conservative form \citep{M71}. However, as we see from Eq.~(\ref{eq:indB}) the generalized Ohm{}\'s law can be easily written in the conservative form and thus any reference to the electric field is unnecessary. The induction Eq.~(\ref{eq:indB}) gives the following jump condition
\bq
\big[B_n\,\left(\v_t + {\vB}_t\right) - \left(v_n + {\vB}_n\right)\,\B_t\big] = 0\,,
\label{eq:bc4}
\eq   
which is identical to the Eq.~(4.7) of Rosenau et al (1979). The boundary conditions, Eq.~(\ref{eq:bc1})-—(\ref{eq:bc4}) together with  
\bq
\big[J_n] = 0\,,
\label{eq:bc5}
\eq 
is the required boundary conditions in the Hall MHD.

\section{dispersion relation}
Following TK65 and SC68 we shall assume that the surface of discontinuity in an incompressible, magnetized planer flow exists across the interface $z = 0$.  The flow velocity $v_x (z)$ is assumed to have the following profile  
\bq
v_x(z) = \left\{
\begin{array}{ll}
v & \mbox{if $z  > 0$}\,,\\
-v & \mbox{if $z < 0$}\,. 
\end{array}
\right.
\label{eq:fpl}
\eq 
The mass density $\rho$ has same value across the interface. An uniform magnetic field $B$ parallel to the $x$ axis is assumed. 

After linearizing and Fourier transforming the equations as
$
\exp\left(\omega\,t + i\,k\,x\right)\,,
$
and applying proper boundary conditions one gets the following dispersion relation [TK65, SC68]
\begin{eqnarray}
\left({\sg_2}^2-{\sg_1}^2\right)^2\,\omega_A^2 + \left(2\,{\omega_A}^2 + {\sg_2}^2 + {\sg_1}^2\right)\,\left(\frac{q_1}{k}\right)\,{\sg_1}^2\times
\nonumber\\
\Big\{
\left({\sg_2}^2+{\omega_A}^2\right)
+ \left(\frac{q_2\,{\sg_2}^2}{q_1\,{\sg_1}^2}\right)\,\left({\sg_1}^2+ {\omega_A}^2\right)
\Big\} = 0\,.
\label{eq:DR}
\end{eqnarray}
Here
\bq 
\sg_j = \omega + i\,k\,v_j\,,\quad\quad \omega_A = k\,\va\,,
\eq
 and 
\bq
\left(\frac{q_j}{k}\right)^2 = 1 + \Big[\frac{\left({\sg_j}^2+ {\omega_A}^2\right)}{\sg_j\,k^2\,\eta_H}\Big]^2\,, 
\label{eq:qbk}
\eq
and $j=1\,,2$.
 Note that the ideal MHD limit corresponds to the absence of Hall term, i.e. $q_j \rightarrow \infty$. 

The above expression for $q_j/k$ is quite complicated and needs to be simplified before making further analytic progress. Since the magnetic field in the Hall MHD evolves under the combined influence of fluid advection and field diffusion ($\v+\vB$), this opens up the possibility of approximating $q_j/k$ in the various limit. For example in the weak diffusive limit, when the fluid advection dominates the field diffusion, 
\bq
\left({\sg_j}^2+ {\omega_A}^2\right) \gtrsim \sg_j\,k^2\,\eta_H\,.
\label{eq:WDL}
\eq
Eq.~(\ref{eq:WDL}) can also be written in the following form
\bq
\left(\frac{\sg_j}{\omega_A}\right) + \left(\frac{\omega_A}{\sg_j}\right) \gtrsim \left(\frac{\omega_A}{\omega_H}\right) \equiv k\,L_H\,,
\eq
which provides the lower bound on the whistler ($\omega_A \ll \sg_j$) and the dressed ion—-cyclotron ($\sg_j \ll \omega_A$) frequencies. In the weak diffusion limit we may approximate $q/k$ as
 \bq
\left(\frac{q_j}{k}\right) \simeq \frac{\left({\sg_j}^2+ {\omega_A}^2\right)}{\sg_j\,k^2\,\eta_H}\,. 
\eq

The dispersion relation Eq.~(\ref{eq:DR}) has been analysed by TK65 and SC68 only in the weak diffusion limit for the super--\alf and \alfc flows. Not only the sub--\alfc flows were not considered in this limit but the other strong diffusion limit was completely ignored. Since it is well known that the Hall diffusion opens a new channel though which the free shear energy can flow to the waves \citep{W99, PW12}, we should anticipate similar outcome also in the present case when the plasma is highly diffusive. 

We shall assume that the shear flow profile is given by Eq.~(\ref{eq:fpl}) and analyse the dispersion relation, Eq.~(\ref{eq:DR}) first in the weak diffusion limit. In this limit    the dispersion relation reduces to the following simple form [SC68] 
\begin{eqnarray}
\left(\omega^2-\omega_0^2\right)\Big[\left(\omega^2-\omega_0^2\right)^2 + 4\,k^2\,v^2\,\omega^2\Big]
\nonumber\\
-4\,\left(k^2\,\eta_H\right)\,{\omega_A}^2\,k^2\,v^2\,\omega = 0\,,
\label{eq:DR1}
\end{eqnarray}
where
\bq
\omega_0^2 = \omega_A^2\left(M_A^2-1\right)\,.
\eq
In the absence of Hall term, Eq.~(\ref{eq:DR1}) gives the usual Kelvin--Helmholtz mode if $M_A^2 > 1$, i.e. $\omega_0^2$ is positive. As noted in SC68, the square bracket term gives two pairs of stable modes. Near the marginal state, when $M_A=1$, the perturbative analysis suggests that the Hall will have destabilizing influence on the equilibrium [SC68].   

We rewrite the above dispersion relation, Eq.~(\ref{eq:DR1}) in the following form 
\begin{eqnarray}
\left(\frac{\omega}{\omega_A}\right)^2
\Bigg[\left(\left(\frac{\omega}{\omega_A}\right)^2+\frac{M_A+3}{2}\right)^2-\frac{5\,M_A^4+8\,M_A^2-3}{4}
\Bigg]
\nonumber\\
-4\,\epsilon\,
\left(k\,L_H\right)M_A^2\,\left(\frac{\omega}{\omega_A}\right)-–\left(M_A^2-1\right)^3=0\,, 
\label{eq:DRx}
\end{eqnarray}
where in order to keep track of the sign of the magnetic field parameter $\epsilon=\pm 1$ has been introduced. In the absence of Hall, the long wavelength  limit, i.e. when $\omega/\omega_A$ becomes large, the above equation gives the usual Kelvin-—Helmholtz mode
\bq
\left(\frac{\omega}{\omega_A}\right)^2 \approx \left(M_A^2-1\right)\,.
\label{eq:kgr}
\eq
Clearly only super--\alfc ($M_A^2 > 1$) waves are unstable in this limit.

\begin{figure}
\includegraphics[scale=0.32]{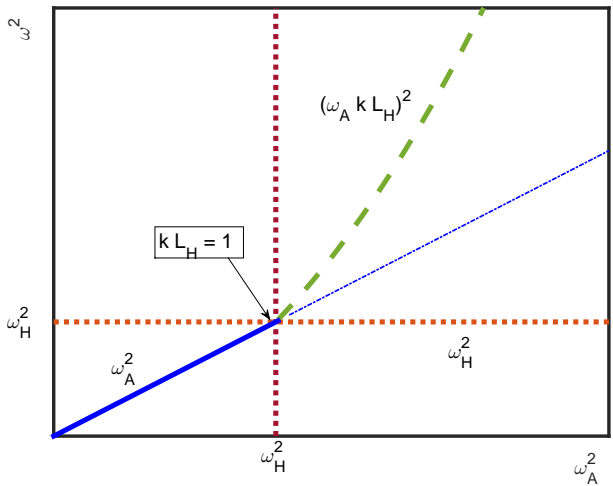}
     \caption{A sketch of the wave propagation in the ideal and Hall MHD is shown in the above figure. The plot $\omega^2$ against $\omega_A^2$ shows various curves corresponding to the \alf, whistler and dressed ion-—cyclotron waves labelled as $\omega_A^2\,,\left(\omega_A\,k\,L_H\right)^2$ and $\omega_H^2$ respectively.}
 \label{fig:fgx}  
\end{figure}

The high frequency whistler ($\omega_A \ll \omega$) 
\bq
\left(\frac{\omega}{\omega_A}\right) = \left(k\,L_H\right)\,,
\label{eq:wlr}
\eq
and the low frequency dressed ion—-cyclotron ($\omega \ll \omega_A$) 
\bq
\left(\frac{\omega}{\omega_A}\right) = \frac{1}{\left(k\,L_H\right)}\,,
\label{eq:icw}
\eq
waves are the normal modes of the Hall MHD \citep{PW08}.  In Fig.~(\ref{fig:fgx}) we sketch the dispersion curves for the ideal and the Hall MHD.  The thick solid line labelled $\omega_A^2$ in the figure corresponds to the normal mode of the ideal MHD $\omega^2 = \omega_A^2$. This line would have continued beyond $k\,L_H =1$ (thin dash—-dot line) in the absence of Hall effect. However, due to the presence of Hall, the whistler [thick dashed line labelled $\left(\omega_A\,k\,L_H\right)^2$] and the dressed ion—-cyclotron (thick dotted line labelled $\omega_H^2$) appears beyond $k\,L_H=1$ in the fourth quadrat of the figure. We see that with increasing $k\,L_H$ whistler becomes the dominant mode in the Hall MHD. To sum, the Hall effect lifts the degeneracy of the ideal MHD by introducing in an otherwise scale free ($k\,L_H = 1$) plasma a scale, namely the Hall scale $L_H$. 

The nature of the whistler and the dressed ion—-cyclotron waves are quite different. While the whistler is caused by the balance between the fluid inertia and the magnetic tension force (like in the MHD), the origin of the dressed ion—-cyclotron wave is electrostatic, i.e. when the magnetic fluctuations are unimportant. Since for the Kelvin-—Helmholtz mode $k\,v \sim \omega$, these two modes operate in two distinct \alf—-Mach parameter space $M_A \gg 1$ and $M_A \ll 1$ respectively. Therefore, the onset of the KH instability in the super and sub--\alfc regions will follow quite distinct path. 

The whistlers have a circular structure \citep{B13}
\bq
\left(\frac{\omega}{k}\right)\delta \B \sim \eta_H\,\curl\delta \B\,,
\eq
and the {\it frozen-—in} condition relates $\delta \B$ to  $\curl \delta \v$ \citep{PW08}. As a result when the circularly polarized whistlers have the same orientation as the fluid vorticity, the instability grows at a higher rate than the corresponding ideal MHD growth rate. 

In the low frequency limit, the dressed ion response time ($\omega^{-1}\sim 1/(k\,v)$) is much faster than the \alf crossing time and thus the magnetic fluctuations are unimportant ($\delta \v/v_A \gg \delta \B/\B$). The dressed ions undergo acoustic type oscillation which in the presence of sub--\alfc flow extract the shear flow energy over $\omega^{-1}$. Since the low frequency ($\omega \ll \omega_A$) limit implies $M_A \ll 1$, balancing the dominant Hall term with the last term in Eq.~(\ref{eq:DRx}) yields 
\bq
\omega \approx \frac{\omega_H}{4\,M_A^2}\,, 
\label{eq:grl}
\eq   
implying that the instability may grow quite rapidly over the dynamical time scale.  Clearly the Hall diffusion of the magnetic field opens up a new channel though which the shear flow energy is fed to the waves.  

When $M_A = 1$, in the absence of the Hall term, the dispersion relation, Eq.~(\ref{eq:DRx}) gives the oscillatory \alf modes $\omega = \pm 2\,i\,\omega_A$. However, if the Hall term is retained in the dispersion relation as a small perturbative correction, the mode becomes unstable with the growth rate [SC68]
\bq
\left(\frac{\omega}{\omega_A}\right) \approx \left(k\,L_H\right)^{1/3}\equiv \left(\frac{k^2\,\eta_H}{\omega_A}\right)^{1/3}\,.
\eq 
Therefore, in the weak diffusion limit irrespective of the nature of the shear flow, the presence of Hall always destabilizes the fluid.  The high frequency whistler (when the electron dynamics is important) is destabilized by the super--\alfc flows and the weakly magnetized ions provides just the neutralizing background. The low frequency ion—-cyclotron wave on the other hand is destabilized by the sub--\alfc shear flows over the ion dynamical timescale.

\begin{figure}
     \includegraphics[scale=0.28]{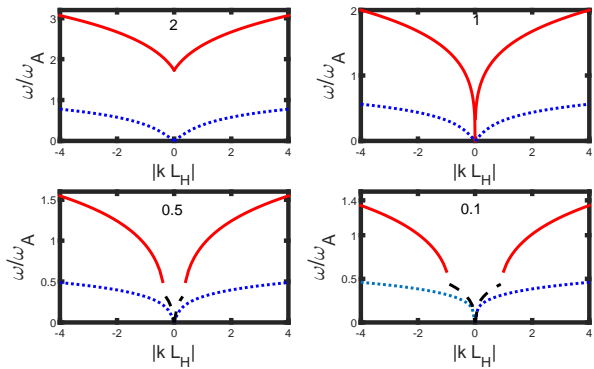}
     \caption{The growth rate (in the units of \alf frequency) against $k\,L_H$ for \alf—-Mach numbers $M_A = 2\,,1\,,0.5\,,0.1$ is shown in the above figure. The solid (dotted) curves correspond to the purely growing (overstable) waves. The mirror symmetric curves with the opposite sign of the Hall term ($\epsilon=-1$) which is tantamount to inverting the sign of the magnetic field (with respect to the $x$ axis) is also shown in the figure.}
\label{fig:fg1}  
\end{figure} 
\begin{figure}
\includegraphics[scale=0.28]{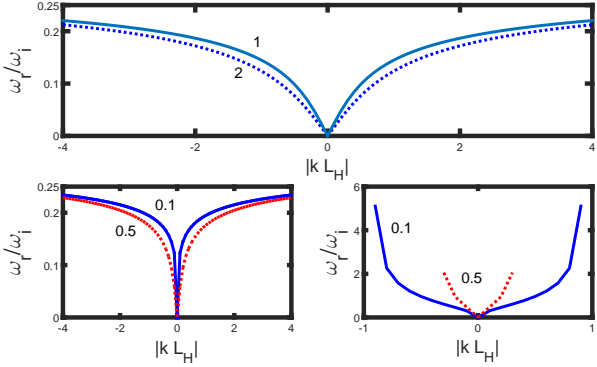}
     \caption{The ratio of the real and imaginary parts, $\omega_r/\omega_i$ against $k\,L_H$ is shown for the overstable modes of Fig.~(\ref{fig:fg1}). The label against the curves are for the various \alf—Mach numbers. The symmetric curves are for the flipped Hall signs.}
\label{fig:fgx2}  
\end{figure} 

We numerically solve the dispersion relation, Eq.~(\ref{eq:DRx}) and plot the result in Fig.~(\ref{fig:fg1}). The growth rate of the KH instability (in the units of \alf frequency $\omega_A$) against $k\,L_H$ for various \alf Mach numbers $M_A=2\,, 1\,,0.5$ and $0.1$ is shown in the figure. We notice that when the flow is super--\alfc, the growth rate of the purely growing Kelvin—-Helmholtz mode increases in the presence of Hall effect. For example the growth rate of the KH instability, which in the MHD flows is $1.73\,\omega_A$ [Eq.~(\ref{eq:kgr})] for $M_A=2$ now becomes $\gtrsim 3\,\omega_A$. In the \alfc ($M_A=1$) and sub--\alfc ($M_A=0.5\,,0.1$) flows, the KH instability is entirely due to the Hall effect albeit this purely growing mode appears only beyond certain $k\,L_H$. Note the gap in the lower panels of the Fig~\ref{fig:fg1}. This suggest that only those fluctuations whose wavelengths are of the order of or, smaller than the Hall scale are Kelvin-—Helmhotz unstable in sub--\alfc flows. The fluctuations of larger wavelengths (smaller $k$) in this case are overstable.

We also notice the presence of less rapidly growing overstable modes (dotted curves) in Fig.~(\ref{fig:fg1}) with the growth rate about $\sim 1/3$ of the purely growing Kelvin—-Helmholtz mode. The presence of this mode is generic in the Hall MHD. For sub--\alfc flows there are two overstable modes. One of the overstable mode appears only when $|k\,L_H|$ is confined within the small range (dashed curve). In its place a purely growing KH mode (solid curve) appears when $|k\,L_H|$ exceeds this range. The other overstable mode (dotted curve) is present at all $|k\,L_H|$.   

The ratio of the real and the imaginary part of the frequencies for $M_A=2\,,1\,,0.5$ and $0.1$ is shown in Fig.~(\ref{fig:fgx2}). In the left lower panel of Fig.~(\ref{fig:fgx2}) this ratio is for the overstable mode appearing at all $|k\,L_H|$ in the sub--\alfc flows. The other overstable mode whose appearance is limited in the neighbourhood of small $|k\,L_H|$, this ratio is shown in the lower right panel of the Fig.~(\ref{fig:fgx2}).  Notice that with increasing $k\,L_H$ (i.e. with  decreasing wavelength) the real part of the frequency increases. This implies that the Hall diffusion channels the shear flow energy more efficiently at the smaller (with respect to the Hall scale) wavelengths than at the longer wavelengths. 

To summarize, we see that (a) the growth rate of the KH instability increases with the increasing $k\,L_H$, i.e. when the Hall dynamics becomes important; (b) the shear flows which are otherwise Kelvin—-Helmholtz stable  without Hall becomes unstable; (c) the sub--\alfc flows also become Kelvin-—Helmholtz unstable in the presence of Hall; and, (d) a new overstable modes whose growth rate is smaller than the purely growing Kelvin—-Helmholtz mode appears in the fluid. The curves in the figure are expectedly mirror symmetric with respect to the sign change of the Hall term in Eq.~(\ref{eq:DRx}).

In order to better appreciate the role of Hall effect and develop a physical understanding of the above results we recast the dispersion relation, Eq.~(\ref{eq:DR1}) in  the following form 
\bq
a\,\left(k\,L_H\right)^6 + b\,\left(k\,L_H\right)^4 
-c\,\left(k\,L_H\right)^2-d = 0\,,
\eq
where
\begin{eqnarray}
a &=& \big[\left(M_A^2-1\right)^4 + 4\left(\frac{\omega}{\omega_H}\right)M_A^2\big]\,,
\nonumber\\
b &=& \big[\left(M_A^2+1\right)^2-4\big]\left(\frac{\omega}{\omega_H}\right)^2\,, 
\nonumber\\
c &=& \left(M_A^2+3\right)\left(\frac{\omega}{\omega_H}\right)^4\,,
\nonumber\\
d &=& \left(\frac{\omega}{\omega_H}\right)^6\,.
\end{eqnarray}
In the low frequency limit, dropping the last term above equation reduces to the following quadratic form 
\bq
a\,\left(k\,L_H\right)^4 + b\,\left(k\,L_H\right)^2-c=0\,.
\eq
The maximum growth rate can be inferred by setting $b^2-4a\,c=0$ which yields
\bq
\left(\frac{\omega_m}{\omega_H}\right) = -\frac{\left(M_A^2-1\right)^2\big[\left(M_A^2+3\right)+4\left(M_A^2-1\right)^2\big]}{16\,M_A^2}\,.
\label{eq:mrt}
\eq
The wavenumber corresponding to the maximum growth rate is
\bq
k_m\,L_H = \frac{\left(M_A^2-1\right)^3\big[\left(M_A^2+3\right)+4\left(M_A^2-1\right)^2\big]}{128\,M_A^4}\,.
\label{eq:wln}
\eq
\begin{figure}
    \includegraphics[scale=0.32]{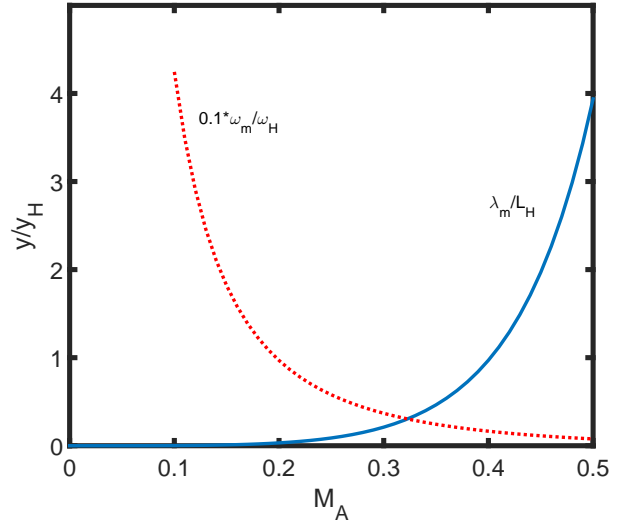}
    \caption{The growth rate (dotted line) and the wavelength (solid curve) against the \alf—-Mach number $M_A$ is shown in the above figure.}
\label{fig:fg3}  
\end{figure} 
The maximum growth rate (dotted line) and the corresponding wavelength (solid line) is plotted against the \alf—-Mach number $M_A$ in Fig.~(\ref{fig:fg3}).  It is clear from the figure that in sub--\alfc flows KH instability is due to the dressed ion--cyclotron wave becoming unstable. Since the dispersion relation Eq.~(\ref{eq:icw}) suggests that these waves are supressed with increasing $k\,L_H$ we see the gradual tapering off of the KH instability with the increasing $\lambda_m/L_H$.  

Note that we may also recast the dispersion relation, Eq.~(\ref{eq:DR1}) as a cubic equation in $M_A^2$ as
\bq
M_A^6 + b_2\, M_A^4 + b_1\, M_A^2-b_0 = 0\,, 
\label{eq:DR2}
\eq
where
\begin{eqnarray}
b_2&=& \left(\frac{\omega}{\omega_A}\right)^2-3\,,
\nonumber\\
b_1&=&4\big[1+\left(\frac{\omega}{\omega_A}\right)\,\left(k\,L_H\right)\big]-\big[\left(\frac{\omega}{\omega_A}\right)^2-1\big]^2\,,
\nonumber\\
b_0&=&\big[1+\left(\frac{\omega}{\omega_A}\right)^2\big]^3\,.
\end{eqnarray}
\begin{figure}
     \includegraphics[scale=0.32]{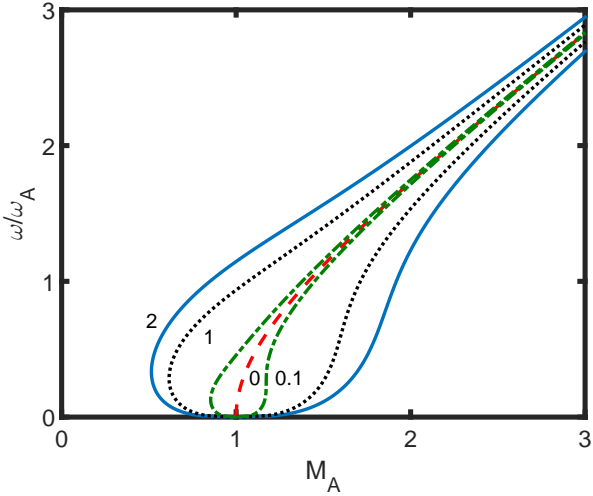}
\caption{The solution of Eq.~(\ref{eq:DR2}) in terms of $M_A$ by varying $\omega/\omega_A$ is plotted when $\B$ is aligned parallel (anti—parallel) to the $x$ axis which corresponds to the positive (negative) sign of the Hall term. The numbers against the curve correspond to the different values of $k\,L_H$. The curve labelled $0$ correspond to the ideal MHD case. The symmetric curves below the label $0$ are for the $\B$ anti—parallel to the $x$ axis which flips the sign of the Hall term.}  
\label{fig:fg4}  
\end{figure}
Although the roots of Eq.~(\ref{eq:DR2}) can be easily written down, we solve this equation numerically by varying $\omega/\omega_A$ for different values of $k\,L_H$. The curve labelled $0$ in Fig.~(\ref{fig:fg4}) correspond to the ideal MHD case while $k\L_H = 0.1\,,1\,, 2$ curves above the label $0$ (below the label $0$) are for the parallel  (with respect to the $x$ axis) and anti—-parallel fields.  The growth rate represented by the label $0$ correspond to the ideal MHD. In this case, only when the flow is super--\alfc i.e. $M_A>1$, the fluid is unstable. 

In the Hall case, both the sub and super--\alfc flows are unstable with the identical growth rate for the parallel and anti—-parallel (to the x-—axis) magnetic field. When $k\,L_H = 1\,,2$, the waves are a mixture of the right circularly polarized whistler and left circularly polarized dressed ion—-cyclotron. In this case, the KH instability is caused in equal measure by both the whistler and the dressed ion-—cyclotron wave growth. However, when $k\,L_H = 0.1$, the waves are predominantly dressed ion—-cyclotron.      

Since, the flow across the $z$ axis is mirror symmetric, the flipping of the background magnetic field sign (and thus the resulting change of the Hall term sign) does not change the orientation of the polarization vector with respect to the flow vortex. Therefore, for both the super and sub--\alfc Mach numbers, the instability grows at the same rate.  In the long wavelength limit, i.e. when $\omega/\omega_A$ becomes large, the growth rate asymptotically approaches ideal MHD limit [Eq.~\ref{eq:kgr}]. This is not surprising given that in the long wavelength limit, the effect of Hall diffusion on the magnetic field tapers off.

\begin{figure}
     \includegraphics[scale=0.28]{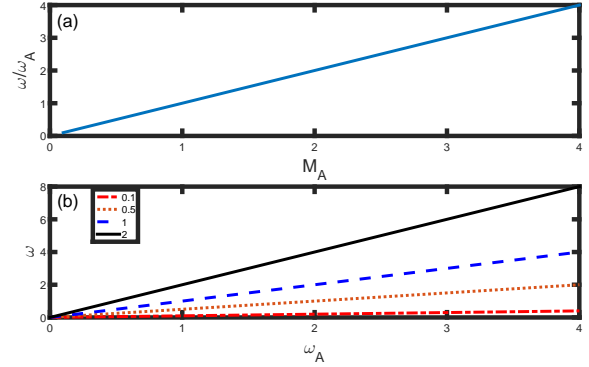}
     \caption{The growth rate $\omega/\omega_A$ in the highly diffusive limit is plotted against the alf—-Mach number $M_A$.}  
\label{fig:fg5}  
\end{figure}

Now we approximate $q_j/k$, Eq.~(\ref{eq:qbk}) in the highly diffusive limit. In this limit 
\bq
\left({\sg_j}^2+ {\omega_A}^2\right) \ll \sg_j\,k^2\,\eta_H\,,
\label{eq:HDL}
\eq
and $q_j/k =1$. Above equation implies that 
\bq
\sg_j \ll k^2\,\eta_H\,,\quad \mbox{and}\,,\quad
\left(\frac{\omega_A^2}{\sg_j}\right) \ll k^2\,\eta_H\,.
\eq 
Note that the long wavelength ($\omega_A\ll \omega_H$) waves are excluded in the highly diffusive limit due to above constraints. Only short wavelength ($\omega_H\ll \omega_A$) fluctuations are permitted in this limit. The dispersion relation Eq.~(\ref{eq:DR}) with $q_j/k=1$ reduces to the following simple form 
\begin{eqnarray}
\left(\frac{\omega}{\omega_A}\right)^6 + \left(M_A^2+2\right) \left(\frac{\omega}{\omega_A}\right)^4-\big[\left(M_A^2+2\right)^2-5\big] \left(\frac{\omega}{\omega_A}\right)^2
\nonumber\\
-–M_A^2\,\left(M_A^2-1\right)^2=0\,, 
\label{eq:DRy}
\end{eqnarray}
 which expectedly gives purely oscillatory modes when the alf-—Mach number is zero, i.e. when there is no free shear energy in the system. However, for any non—-zero \alf--Mach number the waves are unstable. For example when $M_A \ll 1$, the above dispersion relation gives 
\bq
\left(\frac{\omega}{\omega_A}\right) \approx \sqrt{M_A}\,.
\label{eq:SMA}
\eq 
For $M_A=1$, the growth rate of the instability is
\bq
\left(\frac{\omega}{\omega_A}\right) = M_A\,,
\label{eq:MMA}
\eq
and for $M_A \gg 1$ the growth rate becomes
\bq
\left(\frac{\omega}{\omega_A}\right) \approx M_A^{3/2}\,.
\label{eq:LMA}
\eq
Similar trend is seen in the numerical solution of Eq.~(\ref{eq:DRy}) in 
Fig.~\ref{fig:fg5}(a). Expectedly the KH instability grows more slowly in the sub--\alfc flows in comparison with the \alfc and super--\alfc flows. 

In Fig.~\ref{fig:fg5}(b) the growth rate is plotted against $\omega_A$ for $M_A = 0.1\,,0.5\,,1\,,2$. We see that the growth rate increases with increasing $\omega_A$ implying that the high frequency (short wavelength) whistler fluctuations grows much more rapidly than the low frequency ion—-cyclotron fluctuations. This is not surprising given that the most efficient free energy transfer in the Hall dominated plasma occurs at the wavelengths of the order of or, smaller than the Hall scale.  To summarize, in the highly diffusive limit, when the Hall diffusion of the magnetic field acts in tandem with the advection of the magnetic field by the fluid, any presence of shear makes the fluid unstable with the growth rate increasing rapidly with the increasing \alf-—Mach number.   

\section{Applications}
\subsection{Earth{}\'s ionosphere }
\begin{figure}
    \includegraphics[scale=0.28]{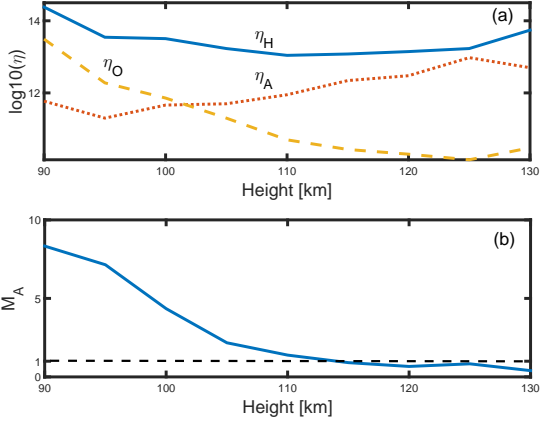}
   \caption{The Hall ($\eta_H [cm^2/s]$), Ohm ($\eta_O $) and ambipolar ($\eta_A$) diffusion coefficients ($cm^2/s$)and the \alf—-Mach number, calculated for the shear flow velocity $100\,\mbox{km}/\mbox{s}$ and $B=0.3\,\mbox{G}$ are plotted against the height in panels (a) and (b) respectively.}
\label{fig:fg7}  
\end{figure}
Large wind shears have been observed in the mesosphere and the lower thermosphere over a wide range of latitude, longitude, seasons and local times for the past five decades \citep{Y10}.  The wind shear often exceeds $70\,\mbox{m}/ \mbox{s} /\mbox{km}$  and can occasionally reach $150\,\mbox{m}/ \mbox{s} / \mbox{km}$.  Presence of the wind shear in the mesosphere and the lower thermosphere may cause the Kelvin--Helmholtz instability which may seed the quasi-—periodic sporadic E structures at the midlatitude \citep{L00}. As can be seen from the Fig.~\ref{fig:fg7}(a), at the lower and mid altitude of the ionosphere the Hall diffusion coefficient [Eq.~(\ref{eq:hdc})] is largest among the three [Ohm, $\eta_O$, ambipolar, $\eta_A$ and Hall, $\eta_H$, \cite{PW08}] non-—ideal MHD diffusion coefficients. The \alf-—Mach number is plotted in the lower panel [Fig.~\ref{fig:fg7}(b)] for the mean shear speed $100\,\mbox{m}/ \mbox{s}/\mbox{km}$ and the magnetic field strength $B = 0.3\,\mbox{G}$. We see that the flow is super--\alfc below $110\,\mbox{km}$ and \alfc or, sub--\alfc beyond $\gtrsim 110\,\mbox{km}$ altitude.  As the sporadic E structures occur at midlatitude, both the super and sub--\alfc flow may cause the KH instability. Thus it is quite plausible that below $110\,\mbox{km}$ the high frequency whistler and above $110\,\mbox{km}$ the low frequency dressed ion—-cyclotron waves may cause Kelvin—-Helmholtz turbulence. 

\begin{figure}
     \includegraphics[scale=0.28]{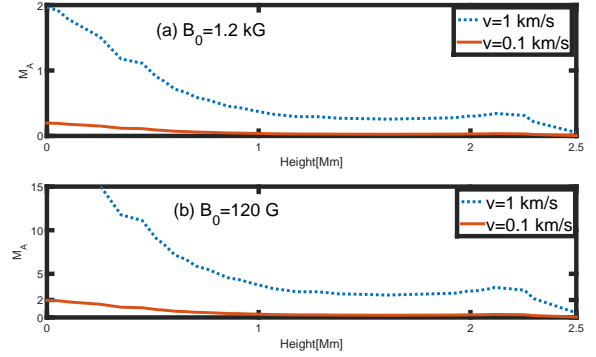}
     \caption{The \alf—-Mach number for the shear flow speed $0.1-1\,\mbox{km}/\mbox{s}$ and the magnetic field stength $B_0=1.2\,\mbox{kG}$ (dashed curve) and $120\,\mbox{G}$ (solid curve) at the base of the  photosphere is plotted against height for the model C of the solar atmosphere \citep{VAL81}.}
\label{fig:fg8}  
\end{figure}
The observed horizontal wavelength of the quasi-—periodic plasma structures at the midlatitude E—-region is typically $\sim 10—-15\,\mbox{km}$ with the period ranging from $\sim 1$ to $10\,\mbox{mins}$ \citep{L00}. Thus the \alf frequency for $\lambda \sim 10\,\mbox{km}$ becomes $\omega_A \sim 0.1-—0.001\,\mbox{s}^{-1}$ and the diffusion frequency $k^2\,\eta_H \sim 4\times 10^2\,\mbox{s}^{-1}$. Here we have assumed $\eta_H = 10^{13}\,\mbox{cm}^2/\mbox{s}$ and $v_A  \sim 10^3—-10^4\,\mbox{cm}/\mbox{s}$. Since the frequency of the quasi-—periodic structures ranges from $\sim 0.0016\,\mbox{s}^{-1}$ to $0.06\,\mbox{s}^{-1}$, the plasma is highly diffusive since Eq.~ (\ref{eq:HDL}) is easily satisfied. Thus assuming $M_A = 2$ between $100-—110\,\mbox{km}$ [Fig.~\ref{fig:fg7}(b)] the growth rate of the KH instability [Eq.~(\ref{eq:LMA})] for $\omega_A = 0.001\,\mbox{s}^{-1}$ is $\sim 0.03\mbox{s}^{-1}$, i.e. about half a minute. Above $110\,\mbox{km}$ flow is slightly sub-\alfc [Fig.~\ref{fig:fg7}(b)], and the instability grows at a rate close to the \alf frequency $0.1\,\mbox{s}^{-1}$, i.e. over a period of $10\,\mbox{s}$. All in all, it is quite plausible that the Hall effect influences the onset of the KH instability at the midlatitude E--region and thus may act as a precursor to the other plasma instabilities. Future numerical simulations with realistic shear flow profiles in the presence of varying magnetic field will allow for the detailed comparison with the observations.   
\subsection{Solar atmosphere}
The upflows of hot plasma in Mm--scale granules and downflows of radiatively cooled plasma in a network of relatively narrow intergranular lanes characterizes the Solar surface convection \citep{N09}.  The magnetic field, which away from the coronal holes appears as the large-scale canopy over the photosphere effectively couples various partially and fully ionized layers of the atmosphere. The upflows and downflows of the plasma in the magnetic flux tubes are subsonic, with the speed of the order of $\sim 0.1—-1\,\mbox{km}/\mbox{s}$ \citep{CI14}. In order to estimate the \alf--Mach number in the photosphere—-chromosphere, we adopt following power law variation of the field with the neutral number density $n_n$  
\bq
B = B_0\,\left(\frac{n_n}{n_0}\right)^{0.3}\,,
\label{eq:scl}
\eq 
where $n_0$ is the number  density of the neutrals at the surface ($h = 0$).  The neutral number density, which varies with the height, is taken from the model C of \cite{VAL81}. Such a field profile captures the essential height dependent features of the observed field in the flux tubes \citep{M97}.    

It is clear from Fig.~\ref{fig:fg8}(a) that for the shear flow speed $\sim 1\,\mbox{km}/\mbox{s}$, as the strong intergranular field ($B_0\sim 1.2\,\mbox{kG}$ at the footpoint) decreases with height, the super--\alfc flow becomes sub--\alfc (dotted curve) in the lower chromosphere ($\sim 700\,\mbox{km}$). However, Hall is the dominant diffusion mechanism only up until approximately this height [Fig.~(1(a), \cite{PW13}] and thus the KH instability will grow at twice the ideal MHD rate due to Hall effect [See top left panel of Fig.~\ref{fig:fg1}]. In the middle and upper chromosphere where the flow is sub--sonic, ambipolar diffusion dominants Hall [Fig.~(1(a), \cite{PW13}]. In this region waves may become unstable due to anisotropic damping by ambipolar diffusion. On the other hand the shear flow speed $\sim 0.1\,\mbox{km}/\mbox{s}$ remains sub--\alfc in the entire photosphere-—chromosphere [solid line in Fig.~\ref{fig:fg8}(b)]. In this case, Hall may drive KH instability in the photosphere and lower chromosphere. Note that even in the sub—-alfc ideal MHD flows, waves may become KH unstable if the magnetic field has a very weak twist, i.e. when additional free energy source is available in the fluid. Note that if the twist in the field is strong, sub--\alfc flows remains stable in the ideal MHD framework \citep{Z10, Z14}. This is because the rolling up of the flow interface (corresponding to the direction of vorticity) is helped by the weak twist in the field; strong twist inhibits this rolling up. Therefore, weak twist in the magnetic field makes sub--\alfc flows in the ideal MHD unstable. However, in the absence of any such twist in the field, only Hall effect, which causes left and right circularly polarized waves, may help the rolling up of the interface. In the present work no additional (to the shear flow) source of free energy is assumed and our focus here is on the role of Hall diffusion only.

In Fig.~\ref{fig:fg8}(b), the \alf-—Mach number, $M_A$ is plotted against height assuming $B_0=120 \,\mbox{G}$ at the footpoint of the flux tube. In this case since Hall is the dominant diffusion in the entire photosphere--chromosphere [Fig.~(1(b), \cite{PW13}], the KH instability will either grow at a faster (than ideal MHD) rate when the flow ($v=1\,\mbox{km}/\mbox{s}$) is super—-\alfc (dotted line)  or, drive the instability in sub—-\alfc (corresponding to $v=0.1\,\mbox{km}/\mbox{s}$) in the middle and upper chromosphere (solid line).   

The Hall scale is of the order of $1-—10\,\mbox{km}$ \citep{PW08, P13} and thus we shall assume that the fluctuation wavelength is of the same order. This gives $k^2\,\eta_H \sim 4\,\mbox{s}^{-1}$ for $\eta_H = 10^{13}\,\mbox{cm}^2/\mbox{s}$. The \alf frequency is $\omega_A \sim 1\,\mbox{s}^{-1}$ for $v_A \sim 0.1\,\mbox{km}/\mbox{s}$. For the dynamical frequency $\omega \lesssim 10^2\,\mbox{s}^{-1}$ [Fig.~2, \citep{P13}], we see that the plasma is in the weak diffusion regime [Eq.~ \ref{eq:WDL}]. Therefore, in the lower photosphere, the growing whistler fluctuations may cause the turbulence and above the upper photosphere the dressed ion-—cyclotron waves may cause the KH turbulence. The growth rate of the this low frequency KH instability [Eq.~\ref{eq:grl}] is quite large. For example for the typical value of $M_A \sim 0.5$ and $\omega_H \sim 10^2\,\mbox{s}^{-1}$ the growth rate is $\sim 10^2\,\mbox{s}^{-1}$. Clearly, the dominance of the Hall diffusion in the lower solar atmosphere may trigger the KH instability both in the upflows and downflows of the plasma along the narrow flux tubes irrespective of whether the flows are sub or, super--\alfc. Recall that the Hall instability may also be caused in the presence of shear flows \citep{PW12, PW13}. However, whereas the Hall instability requires the presence of a transverse (to the magnetic field) shear flow, the present investigation of the KH instability is for the parallel/antiparallel field--flow geometry.       

\subsection{Protoplanetary Discs}
It is not well understood how the planetisimals--the kilometre--sized precursors of the real planets formed. It is believed that when the thin dust layer (which settles to the midplane due to the gravity of the protosolar nebula) in a protoplanetary disc becomes gravitationally unstable, planetisimals forms \citep{S69, GW73}. However, as the dust settles to the midplane the shear due to the Keplerian rotation of the disc increases and eventually the layer becomes Kelvin--Helmholtz unstable. The resulting turbulence in the disc will diffuse the dust away from the midplane preventing its density from reaching the critical value required for the gravitational instability. Thus shear induced turbulence may prevent the formation of planetisimals \citep{W80, W84, S98}. The midplane of a protoplanetary disc is often highly diffusive and the magnetic field may be poorly coupled to the field. The Hall diffusion could be important at the disc midplane only between $3-5$ pressure scale heights for a mG field at $1$ AU [Figs.~5 and 7, \cite{W07}] while at $5$ AU it becomes important much closer ($\gtrsim 1$ scale height) to the disc midplane [Figs.~10 and 11, \cite{W07}]. For stronger magnetic field $B\sim 0.1-0.01\,\mbox{G}$, Hall dominates all other non-—ideal MHD diffusive processes at $5$ AU near the disc midplane. Assuming the gas density distribution for a minimum mass solar nebula 
\bq
\rho_g=2.8\times10^{-11}\,\left(\frac{R}{5\,\mbox{AU}}\right)^{-11/4}\,g\,\mbox{cm}^{-3}\,,
\eq
one gets for $B=0.1\,\mbox{G}\,,$ $v_A \simeq 0.1\,\mbox{km}/\mbox{s}$. Equating the disc scale height
\bq
h = 0.35\,\left(\frac{R}{5\,\mbox{AU}}\right)^{5/4}\,\mbox{AU}\,,
\eq   
with the $k^{-1}$, we get the \alf frequency $\omega_A\sim 10^{-7}\,\mbox{s}^{-1}$.  For typical value of $\eta_H \sim 10^{14}\,\mbox{cm}^2/\mbox{s}$ [Figs.~4, \cite{W07}] and assuming that the $\omega$ is of the order of the disc dynamical frequency $\Omega_K \sim 10^{-8}\,\mbox{s}^{-1}$, we see that the disc is highly diffusive as  Eq.~(\ref{eq:HDL}) is easily satisfied. Thus the KH instability growth rate in the disc depends on the \alf—-Mach number. For example when the flow is \alfc, the instability growth rate is $\omega = \omega_A \sim 10^{-7}\,\mbox{s}^{-1}$. Therefore, over the disc dynamical time scale, the Hall diffusion, irrespective of the nature of the shear flow, may destabilize the disc. The ensuing turbulent mixing of the material could further slowdown the planet formation.

\subsection{Molecular Clouds}
Molecular clouds are the sites of star formation. It is believed that the star formation is associated with an intrinsic scale over which the gravitational collapse of massive cloud complex occurs \citep{L95}. This intrinsic scale possibly depicts the size of the molecular cloud cores of diameter $\sim 0.1\,\mbox{pc}$, the ambient temperature $10\,\mbox{K}$, number density $\sim 10^{4}\,\mbox{cm}^{-3}$ and line width $\sim 0.5\,\mbox{km}/\mbox{s}$ \citep{M94}. Since the cloud core is embedded in the ambient molecular gas, it may suffer KH instability due to its relative motion with the surrounding. The timescale of the hydrodynamic KH instability is much smaller ($0.05\,\mbox{Myr}$) than the typical cloud lifetime which is several tens of megayear \citep{K96}.

The presence of mG field over several hundred AU in the high density ($n_n \sim 10^8-10^{10}\,\mbox{cm}^{-3}$) region is inferred from the Zeeman splitting of OH and $\mbox{H}_2\,\mbox{O}$ maser lines \citep{H87}. The millimetre and far-infrared observations indicate the presence of large scale ($0.1–-10\,\mbox{pc}$) ordered fields \citep{THH95}. Although the fractional ionization of the cloud is quite low ($\sim 10^{-4}-—10^{-8}$), the observed persistence of the turbulent motion in the cloud is often attributed to the presence of MHD waves \citep{AM75}. However, since the matter is largely neutral the magnetic field  may diffuse through the cloud. The profiles of various diffusivities in the cloud \citep{WN99, W16} suggests that the Hall dominates ambipolar and Ohm when the gas density is $\sim 10^{11}—-10^{14} \mbox{cm}^3$. Taking $\eta_H \sim 10^{20}\,\mbox{cm}^3$ [Fig.~1, \cite{W16}], and the shear flow velocity (between the core and the gas) $ v\sim 1\,\mbox{km}/\mbox{s}$ \citep{K96} we get the Hall scale $L_H \sim 100\,\mbox{AU}$ for trans--\alfc flows $M_A \sim 1$. Thus assuming $k \sim 1/L_H$, the KH growth rate in the highly diffusive limit is $\omega = \omega_A \sim 10^{-10}\,\mbox{s}^{-1}$ which translates to $t_{\mbox{KH}} \sim 10^{-3}\,\mbox{Myr}$. Note that this timescale is much faster than the hydrodynamic timescale $ t_{\mbox{KH}} \sim 0.05\,\mbox{Myr}$ and much much smaller than the typical cloud lifetime. Clearly, the onset and the collapse of the molecular cloud cores will be affected by the presence of the Hall effect so long as it is not mitigated by other diffusive processes. 

Note that the above analysis completely neglects the presence of grains in the cloud. 
In dense cloud cores, charged grains are more numerous than the plasma particles and its presence not only affects the ionization structure of the cloud but also its gas phase abundances \citep{H97, WN99, W07}. Owing to the low ionization fraction, grains in the cloud are either neutral or carry $\pm 1-\pm 2$ electronic charge \citep{N80, NNU91, WN99, W07}. Further, owing to the large mass and size distribution, the grains can couple directly as well as indirectly to the magnetic field \citep{CM93}. Therefore, the magnetic diffusion of the cloud will be severely affected by the grains.

\section{Summary}
This paper investigates the Kelvin-—Helmholtz instability in the Hall MHD framework. The analysis has been carried out in two limiting cases: (a) weak diffusion limit, when Hall diffusion of the field is weak and, (b) strong diffusion limit, when  Hall   diffusion is the main cause of the field evolution. Following is the summary of the result.

{\bf In the weak diffusion limit:}\\

1. Both the super and sub--\alfc shear flows are unstable due to the presence of Hall effect although the nature of the instability is quite different in two cases.\\  
2. The high frequency whistler waves are unstable in the presence of super--\alfc shear flows. The growth rate of the instability increases with increasing $k\,L_H$\\ 
3. The low frequency ion--cyclotron waves becomes unstable in the presence of sub--\alfc shear flows. The growth rate of the instability decreases with increasing $k\,L_H$\\  
4. A new overstable modes whose growth rate is smaller than the purely growing Kelvin—-Helmholtz mode also appears in the fluid.\\ 

{\bf In the strong diffusion limit:}

5. The growth rate of the instability depends only on the alf-—Mach number and is independent of the Hall diffusion coefficient.\\
6. The growth rate linearly increase with the \alf frequency and have a smaller (in comparison to the \alfc or, super--\alfc flows) slope for the sub--\alfc flows. Thus the sub--\alfc flows grows at a slower rate than the super--\alfc flows.  

\section*{Acknowledgments}
It is my great pleasure to thank Prof. Mark Wardle for encouraging to investigate this problem and countless discussion on this subject. This research has made use of NASA's Astrophysics Data System.

\end{document}